\documentstyle[12pt]{article}

\begin{document}

\title{The leader operators of the $(d+1)$-dimensional relativistic 
rotating oscillators}
\author{Ion I. Cot\u aescu, Ion I. Cot\u aescu Jr. and Nicolina Pop \\
{\em West University of Timisoara}\\
{\em V. P\^ arvan Ave. 4, RO-300223 Timi\c soara, Romania}}

\maketitle

\begin{abstract}

The main pairs of leader operators of the quantum models of relativistic 
rotating oscillators in arbitrary dimensions are derived. To this end one 
exploits the fact that these models generate P\"{o}schl-Teller radial problems 
with remarkable properties of supersymmetry and shape invariance. 

Pacs: 04. 62. +v
\end{abstract}


\newpage

\section{Introduction}

In general relativity one can build  geometric models of pure harmonic or 
even rotating oscillators constituted by test particles freely moving in 
suitable curved spacetimes. In these models the classical and quantum geodesic 
motions represent either harmonic oscillations in $(1+1)$ dimensions 
\cite{C1}-\cite{C4} or oscillations that can be accompanied by supplemental 
uniform rotations in three space dimensions \cite{C1R} as well as in the 
general case of any $(1+d)$ dimensions \cite{C5,C6}.

The backgrounds of these models have static central charts with 
pure or deformed anti-de Sitter metrics. For $d>1$ all the models are of 
rotating oscillations apart from the anti-de Sitter ones where the rotation 
effects disappear \cite{C1R,C5}. In these central charts, the Klein-Gordon 
equations, describing the quantum scalar test particle, can be analytically 
solved leading to standard P\"{o}schl-Teller radial problems \cite{PT,NS}.
This means that these models can be studied  using the method of supersymmetry 
and shape invariance \cite{DKS} with minimal changes requested by the specific 
form of the Klein-Gordon equation \cite{C3,C6}. In this way we have shown that 
all the (1+1)-dimensional P\"{o}schl-Teller models have the same dynamical 
algebra,  $so(1, 2)$, formed by a pair of leader operators and the operator of 
the number of oscillation quanta \cite{C2,C3,C4}. In the case of any 
dimensions we have studied the supersymmetry of the relativistic 
P\" oschl-Teller radial problems but without to write down the leader 
operators \cite{C6}.    
 
In the present letter we should like to continue the study of the
supersymmetry of the radial  relativistic P\"{o}schl-Teller  problems in
arbitrary dimensions. Our purpose is to derive the principal pairs 
of leader operators able to increase or decrease the quantum numbers of  
the radial modes of these models. 

In section 2 we briefly present the radial problems of the models of (rotating)
oscillators in $(d+1)$ dimensions showing that these are P\"{o}schl-Teller 
problems with a very simple parametrization since all the physical parameters 
concentrate into only one parameter which characterize the model. 
The section 3 is devoted to the supersymmetry and the shape 
invariance of the radial potentials that lead to the Rodrigues formula 
of the normalized radial functions and some useful identities given in the 
next section. These results help us to identify in section 5 the principal 
pairs of leader operators, including a pair which change with $\pm 1$ the value 
of the parameter of these models. The final results are presented in the last 
section. 
 
\section{The radial P\"{o}schl-Teller problems}

Let us consider the models of relativistic 
rotating oscillators of the de Sitter type \cite{C6} with the static central 
charts of coordinates $(t, {\bf x})$, where $t=x^0$ is the time while   
${\bf x}=(x^1,  x^2, . . . ,  x^d)$ represent the Cartesian space coordinates.
In these models the oscillating test particle is described by a scalar quantum 
field $\Phi$ of mass $M$,  minimally coupled with the gravitational field.  
The quantum scalar modes are given by the particular solutions of the 
Klein-Gordon equation  
\begin{equation}\label{3}
\frac{1}{\sqrt g}\partial _\mu (\sqrt g g^{\mu \nu }\partial _{\nu }\Phi )
+M^2\Phi =0, \quad g=\left\vert \det (g_{\mu \nu })\right\vert\,.
\end{equation}

The spherical variables can be separated  using generalized spherical
harmonics, $Y_{l(\lambda )}^{d-1}{({\bf x}/r)}$,  where $r=\left \vert {\bf x}%
\right \vert$.  These are normalized eigenfunctions of the angular Laplace
operator \cite{T},
\begin{equation}\label{4}
-\Delta _{s}{Y}_{l(\lambda )}^{d-1}{({\bf x}/r)=l(l+d-2)Y}_ {l(\lambda
)}^{d-1}{({\bf x}/r)}\,,  
\end{equation}
corresponding to eigenvalues depending on the angular quantum number $l$
which takes only integer values,  $l=0, 1, 2,...$,  selected by the boundary
conditions on the sphere $S^{d-1}$ \cite{T}. The particular solutions of 
energy $E$ (and positive frequency) \cite{C6}, 
\begin{equation}
{\Phi }_{E, l(\lambda )}^{(+)}{(t, {\bf x})=}\frac{1}{\sqrt{2E}}{(\omega }\cot 
{\ \omega r)}^{\frac{d-1}{2}}R_{E,l}(r)Y_{l(\lambda )}^{d-1}({\bf x}/r)e^{-iEt}
\,,\label{5}
\end{equation}
involve the radial wave function $R_{E,l}\propto R_{k,l,n_r}$ which satisfy the 
radial equation 
\begin{equation}\label{6}
\left [ -\frac{1}{\omega ^{2}}\frac{d^{2}}{dr^{2}}+\frac{2s(2s-1)}{\sin
^{2}\omega r}+\frac{2p(2p-1)}{\cos ^{2}\omega r}\right ] R_{k, l, n_{r}}=\nu
^{2}R_{k, l, n_{r}}\,,  
\end{equation}
where $\nu$ depends on $E$ as in Ref. \cite{C6} and we have
\begin{equation}
k=\sqrt{\frac{M^{2}}{\omega ^{2}}+\frac{d^{2}}{4}}+\frac{d}{2},  \quad
2s=l+\delta,  \quad 2p=k-\delta \,, \quad \delta =\frac{d-1}{2}\,.  \label{7}
\end{equation}
In this  parametrization the specific parameter $k$ concentrates all the other 
ones, playing thus the role of the principal parameter of our models. For this 
reason we denote the model by $[k]$ understanding that it generates the radial 
problems $(k,l)$ with different values of $l$, each one having its own sequence of 
radial wave functions, $R_{k, l, n_{r}}(r) $. These are  labeled by the radial 
quantum number $n_{r}=0, 1, 2, . . . $  giving the quantization condition 
$\nu =2(n_{r}+s+p)$ \cite{C5,C6}. In this way one obtains  a typical 
P\"{o}schl-Teller problem with square integrable radial functions with respect 
to the radial scalar product 
\begin{equation}\label{sc}
\langle R, R^{\prime }\rangle =\int_{D_{r}}dr\,R^{\ast }(r)R^{\prime }(r)\,.
\end{equation}
 
\section{The radial supersymmetry}

We have shown \cite{C6} that the supersymmetric formalism of the radial 
problems $(k, l)$ can be constructed in the same manner as in the case of 
the relativistic $(1+1)$-dimensional P\"{o}schl-Teller models \cite{C3}. The 
first step is to introduce the operator 
\begin{equation}  \label{18}
\{\Delta [V]R\}(r)=\left[-\frac{1}{\omega^2}\frac{d^{2}}{dr^{2}}+V(r)\right]
R(r)\,, 
\end{equation} 
which should play the same role as the Hamiltonian of the one-dimensional 
nonrelativistic problems. Furthermore, one observes that from Eq. (\ref{4}) we 
can derive the normalized ground-state radial functions 
\begin{equation}\label{19}
R_{k, l, 0}(r)=\sqrt{2\omega }\left [\frac{\Gamma (k+l+1)}{\Gamma (l+\frac{d}{2%
} )\Gamma (k+1-\frac{d}{2})}\right ]^{\frac{1}{2}}\sin ^{2s}{\omega r}\cos
^{2p}{\ \omega r}\,,  
\end{equation}
producing the radial superpotentials  
\begin{equation}\label{10}  
W_{k,l}( r)=-\frac{1}{\omega\, R_{k, l, 0}(r)}\frac{dR_{k, l, 0}(r)}{dr}
=2p\tan \omega r-2s \cot \omega r\,.
\end{equation}
In its turn, each superpotential $W_{k,l}$ give rise to the pair of 
superpartner radial potentials,  
\begin{eqnarray}  
{V}_{\pm } (k, l, r)&=&\pm \frac{dW_{k,l}( r)}{dr}{+W_{k,l}( r)}^{2}
\nonumber\\
&=&\frac{2s(2s\pm 1)}{\sin ^{2}\omega r}{+}\frac{2p(2p\pm 1)}{\cos
^{2} \omega r}-(2s+2p)^{2}\,.\label{21}
\end{eqnarray}
Now Eq. (\ref{4}) can be rewritten as 
\begin{equation}\label{15}
{\Delta \lbrack V}_{-}{(k, l)]R}_{k, l, n_{r}}{=d}_{k, l, n_{r}}
{R}_{k, l, n_{r}}\,,
\end{equation}
where 
\begin{equation}\label{16}
{d}_{k, l, n_{r}}=4 {n}_{r}{(n}_{r}{+k+l)}\,.  
\end{equation}

According to the standard procedure \cite{DKS,C6}, we define the pair of 
adjoint operators,  $A_{k, l}$ and $A_{k, l}^{\dagger }$,  having the action 
\begin{eqnarray}
(A_{k, l}R)(r)&=&\left[ \frac{1}{\omega}\frac{d}{dr}+W_{k,l}( r)\right] R(r)\,,
\label{A01}\\ 
(A_{k, l}^{\dagger }R)(r)&=&\left[-\frac{1}{\omega}\frac{d}{dr}
+W_{k,l}( r)\right] R(r)\,.\label{A02}   
\end{eqnarray}
These are the supersymmetry generators that allow us to write 
\begin{equation}\label{DDAA}
\Delta [V_{-}(k, l)]= A_{k, l}^{\dagger }A_{k, l}\,,
\qquad \Delta [  V_{+}(k, l)]=A_{k, l}A_{k, l}^{\dagger }\,.   
\end{equation}
The crucial point of this approach is that the radial potentials $V_{-}(k, l)$ 
and $V_{+}(k, l)$ are shape invariant since 
\begin{equation}\label{27}
{V}_{+}(k, l, r)=V_{-}(k+1, l+1, r)+4(k+l+1)\,.  
\end{equation}
Consequently,  we can verify that 
\begin{equation}\label{28}
\Delta [V_{+}(k, l)]R_{k+1, l+1, n_{r}-1}=d_{k, l, n_{r}}R_{k+1, l+1, n_{r}-1}
\,,\quad n_{r}=1, 2, . . . \,,
\end{equation}
which means that the spectrum of the operator $\Delta [V_{+}(k, l)]$
coincides with that of $\Delta [ V_{-}(k, l)]$,  apart from the lowest
eigenvalue $d_{k, l, 0}=0$.  From Eqs. (\ref{DDAA}) combined with Eq. (\ref{15}) 
it results that the normalized radial functions satisfy 
\begin{eqnarray}
A_{k, l}{R}_{k, l, n_{r}}&=&\sqrt{d_{k, l, n_{r}}}R_{k+1, l+1, n_{r}-1}\,,\\
{A}_{k, l}^{\dagger }{R}_{k+1, l+1, n_{r}-1}&=&\sqrt{d_{k, l, n_{r}}}
{R}_{k, l, n_{r}}\,. 
\end{eqnarray}
Hence, the superpartner radial problems given  by the superpartner radial 
potentials ${V}_{\pm }{\ (k, l, r)}$ are $(k, l)$ and, respectively, 
$(k+1, l+1)$. Since both the parameters of these radial 
problems increase simultaneously with one unit, we can say that the parameter 
$k$ simulates the behavior of a quantum number playing a similar role as 
the orbital quantum number $l$. 

\section{The normalized radial functions}

The above presented properties lead to the operator version of the Rodrigues 
formula of the normalized radial functions \cite{C6},  
\begin{eqnarray}
R_{k,l,n_{r}}&=&\frac{1}{2^{n_{r}}}\left[\frac{\Gamma(n_{r}+k+l)}
{n_{r}!\,\Gamma(2n_{r}+
k+l)}\right]^{\frac{1}{2}}{A}_{k,l}^{\dagger}{A}_{k+1,l+1}^{\dagger}
\cdots\nonumber\\
&&\cdots {A}_{k+n_{r}-1,l+n_{r}-1}^{\dagger} R_{k+n_{r},l+n_{r},0}\,.
\label{Rod}
\end{eqnarray}
This expression can be brought in the usual standard form observing that in 
the new variable $u=\cos 2\omega r$ the operators (\ref{A01}) and (\ref{A02}) 
read 
\begin{eqnarray}
A_{k,l}&=&-2(1-u)^{s+\frac{1}{2}}(1+u)^{p+\frac{1}{2}}
\frac{d}{du}(1-u)^{-s}(1+u)^{-p}\,,\label{A1}\\
A_{k,l}^{\dagger}&=&2(1-u)^{-s+\frac{1}{2}}(1+u)^{-p+\frac{1}{2}}\frac{d}{du}
(1-u)^{s}(1+u)^{p}\,.\label{A2}
\end{eqnarray}
Then Eq.(\ref{Rod}) yields the following normalized radial functions: 
\begin{equation}\label{22}
{R}_{k, l, n_{r}}{(u)=N}_{k, l, n_{r}}{\chi }_{_{{k, l, n}_{r}}}{(u)}\,,
\end{equation}
where 
\begin{equation}\label{23}
{N}_{k, l, n_{r}}=\frac{1}{2^{n_r+\frac{k+l}{2}}}\left [ 
\frac{ 2\omega (2n_{r}+k+l)\Gamma (n_{r}+k+l)}{n_{r}!\Gamma (n_{r}+l+
\frac{d}{2} )\Gamma (n_{r}+k+1-\frac{d}{2})} \right ]^{\frac{1}{2}}\,,  
\end{equation}
\begin{eqnarray}
{\chi }_{_{{k, l, n}_{r}}}(u)\ &=&(1-u)^{-s+\frac{1}{2}}{(1+u)}^{-p
+\frac{1}{2}}\nonumber\\
&&\times \frac{d^{n_{r}}}{du^{n_{r}}}{(1-u)}^{2s +n_{r}-
\frac{1}{2}}{(1+u)}^{2p +n_{r}-\frac{1}{2}}\,.\label{chi}  
\end{eqnarray}

This expression is suitable for deriving some basic identities that will 
help us to write down the action of the leader operators.
First, we consider  the commutation relation 
\begin{equation}  \label{40}
\left[1-u^2\,,\frac{d^{n+1}}{du^{n+1}}\right]=2u(n+1) 
\frac{d^{n}}{du^{n}}+n(n+1)\frac{d^{n-1}}{du^{n-1}}\,.
\end{equation}
and using Eq. (\ref{chi}) we obtain the basic identity 
\begin{eqnarray}
(1-u^{2})\frac{d{\chi }_{k, l, n_{r}}}{du}&=&
\left[u\left(2n_{n_r}+1+\frac{k+l}{2}\right)+ 
\frac{k-l}{2}+\delta\right] \chi_{k, l, n_{r}}\nonumber\\
&&+\chi_{k, l, n_{r}+1} 
+n_{r}(n_{r}+1)\sqrt{1-u^{2}}{\chi}_{k+1, l+1, n_{r}-1}  \label{41}
\end{eqnarray}
Furthermore, from Eq. (\ref{chi}) we deduce other two useful 
auxiliary identities,
\begin{eqnarray}
&&\chi_{k,l,n_r+1}+2(n_r+l+\delta+\textstyle\frac{1}{2})\chi_{k,l,n_r}
\nonumber\\
&&~~~~~~~~~~~~~~~-(2n_r+k+l+1)\sqrt{1-u}\, \chi_{k,l+1,n_r}=0\,,\label{idl}\\ 
&&\chi_{k,l,n_r+1}-2(n_r+k-\delta+\textstyle\frac{1}{2})\chi_{k,l,n_r}
\nonumber\\
&&~~~~~~~~~~~~~~~+(2n_r+k+l+1)\sqrt{1+u}\, \chi_{k+1,l,n_r}=0\,,\label{idk} 
\end{eqnarray}
which will help us to calculate the action of the first order operators.

\section{Pairs of leader operators}

The next step is to identify the pairs of leader operators shifting not only 
the genuine quantum numbers $l$ and $n_r$ but also the parameter $k$ which 
simulates the behavior of a quantum number. To this end it is convenient to 
work only with the functions $\chi$ giving up the normalized factors the 
presence of which could lead to some useless difficulties.

We observe that the action of the supersymmetry generators on the functions 
$\chi_{k,l,n_r}$ reads
\begin{eqnarray}
\textstyle\frac{1}{2}
A_{k,l}\,\chi_{k,l,n_r}&=&n_r(n_r+k+l)\,\chi_{k+1,l+1,n_r-1}\,,\label{Achi1}\\ 
\textstyle\frac{1}{2}
A_{k,l}^{\dagger}\,\chi_{k+1,l+1,n_r-1}&=&\chi_{k,l,n_r}\,, \label{Achi2}
\end{eqnarray}
and we look for other pairs of operators with similar effects.
Assuming that the pairs of leader operators depend linearly on 
the supersymmetry generators, we define the  general form of the leader 
operators as  
\begin{eqnarray}
\tilde A^{(-)}_{k,l,n_r}&=&\sqrt{\phi}\left(\frac{1}{2}A_{k,l}+H_{k,l,n_r}
\right) \,,\\ 
\tilde A^{(+)}_{k,l,n_r}&=&\phi\left(\textstyle\frac{1}{2}A^{\dagger}_{k,l}+
H_{k,l,n_r}\right)\frac{1}{\sqrt{\phi}}\,,
\end{eqnarray}
where $H_{k,l,n_r}$ and $\phi$ are functions of $u$. These operators may have 
different actions 
\begin{eqnarray}
\tilde A^{(-)}_{k,l,n_r}\chi_{k,l,n_r}&=&c^{(-)}_{k,l,n_r}
\chi_{k+\Delta k,l+\Delta l, n_r+\Delta n_r}\,,\\ 
\tilde A^{(+)}_{k,l,n_r}\chi_{k+\Delta k,l+\Delta l, n_r+\Delta n_r}&=&
c^{(+)}_{k,l,n_r}\chi_{k,l,n_r}\,,
\end{eqnarray}
depending on the values of the shifts $\Delta k$, $\Delta l$ and $\Delta n_r$ 
we choose. In addition, any pair of such operators must obeys the identity 
\begin{eqnarray}
&&\tilde A^{(+)}_{k,l,n_r}\tilde A^{(-)}_{k,l,n_r}\,\chi_{k,l,n_r}=
\phi \left[n_r(n_r+k+l)+H_{k,l,n_r}^2\right.\nonumber\\
&&\left.+\sqrt{1-u^2}\,\frac{dH_{k,l,n_r}}{du}+W_{k,l}H_{k,l,n_r}
\right]\chi_{k,l,n_r}
=c^{(+)}_{k,l,n_r}c^{(-)}_{k,l,n_r}\chi_{k,l,n_r}\,,\label{AAcc}
\end{eqnarray} 
resulted from Eqs. (\ref{15}), (\ref{16}) and (\ref{DDAA}).

Thus we have formulated a problem that may be solved finding the concrete 
form of the functions $H_{k,l,n_r}$ and $\phi$. First of all, we observe  
that some previous results \cite{NS,B} suggest us to consider functions of 
the form
\begin{equation}\label{HH}
H_{k,l,n_r}(u)=\alpha_{k,l,n_r}\sqrt{\frac{1-u}{1+u}}
+\beta_{k,l,n_r}\sqrt{\frac{1+u}{1-u}}\,.
\end{equation}
Furthermore,  we have to use the Eqs. (\ref{A1}), (\ref{A2}) and (\ref{AAcc}) 
for finding the constants $\alpha_{k,l,n_r}$, $\beta_{k,l,n_r}$ and 
$c_{k,l,n_r}^{(\pm)}$. The shifting action will be deduced from Eqs. 
(\ref{Achi1}), (\ref{Achi2}) combined with the identities  
(\ref{41}), (\ref{idl}) and (\ref{idk}). 

In this way we find many solutions which will be labeled by an index preceding 
the symbols. The first three pairs of operators, 
$_m\tilde A^{(\pm)}_{k,l,n_r}$, with $m=1,2,3$,  have their defining elements 
as given in the table below 
\begin{center}
\begin{tabular}{ccccclll}
Nr.&symbol&$\Delta k$&$\Delta l$&$\Delta n_r$&$\phi(u)$&$\alpha_{k,l,n_r}$&
$\beta_{k,l,n_r}$\\
&&&&&&\\
{ 1}&$_1\tilde A^{(\pm)}_{k,l,n_r}$&1&0&0&${1-u}$&0&$n_r+k+l$\\
{ 2}&$_2\tilde A^{(\pm)}_{k,l,n_r}$&0&1&0&${1+u}$&$n_r+k+l$&0\\
{ 3}&$_3\tilde A^{(\pm)}_{k,l,n_r}$&0&1&-1&${1+u}$&$n_r$&0\\
\end{tabular}
\end{center}
while their action is defined by the constants listed in the next table
\begin{center}
\begin{tabular}{cll}
Nr.&$c^{(-)}_{k,l,n_r}$&$c^{(+)}_{k,l,n_r}$\\
&&\\
1&$n_r+k+l$&$2(n_r+k-\frac{1}{2}d+1)$\\
2&$n_r+k+l$&$2(n_r+l+\frac{1}{2}d)$\\
3&$2n_r(n_r+k-\frac{1}{2}d)$&1\\
\end{tabular}
\end{center}
In addition, there is a fourth solution which is more complicated corresponding 
to $\Delta k=\Delta l=0$ and $\Delta n_r=1$. In this case we find that 
the pair of operators $_4\tilde A^{(\pm)}_{k,l,n_r}$ is defined by 
$\phi(u)=1-u^2$ and the function $_4H_{k,l,n_r}$ of the form (\ref{HH}) with 
the coefficients 
\begin{equation}
_4\alpha_{k,l,n_r}=\frac{n_r(n_r+l+\frac{1}{2}d-1)}{2n_r+k+l-1}\,, \quad
_4\beta_{k,l,n_r}=-\frac{n_r(n_r+k-\frac{1}{2}d)}{2n_r+k+l-1}\,. 
\end{equation}
The respective action constants read 
\begin{eqnarray}
_4c^{(-)}_{k,l,n_r}&=&\frac{4n_r(n_r+k-\frac{1}{2}d)(n_r+l+\frac{1}{2}d-1)}
{2n_r+k+l-1}\,, \\
_4c^{(+)}_{k,l,n_r}&=&\frac{n_r+k+l-1}{2n_r+k+l-1}\,. 
\end{eqnarray}

We note that for each pair producing the shifts 
$\Delta k,\,\Delta l,\, \Delta n_r$ one can define another pair performing 
the inverse shifts $-\Delta k,\,-\Delta l,\, -\Delta n_r$. As a matter of 
fact this means to change among themselves the roles of the operators 
$\tilde A^{(-)}$ and $\tilde A^{(+)}$ in a suitable  parametrization. 

\section{Final results}

Using the above results we can write the action of the pairs of leader 
operators upon the normalized wave functions $R_{k,l,n_r}$ as
\begin{eqnarray}
_m\tilde A^{(-)}_{k,l,n_r}R_{k,l,n_r}&=&_mC^{(-)}_{k,l,n_r}
R_{k+\Delta k,l+\Delta l, n_r +\Delta n_r}\,,\\
_m\tilde A^{(+)}_{k,l,n_r}R_{k+\Delta k,l+\Delta l, n_r +\Delta n_r}&=&
_mC^{(+)}_{k,l,n_r}R_{k,l,n_r}\,.
\end{eqnarray}
where
\begin{equation}
_mC^{(\pm)}_{k,l,n_r} = 
{_mc}^{(\pm)}_{k,l,n_r}\left[\frac{N_{k+\Delta k,l+\Delta l, 
n_r +\Delta n_r}}{N_{k,l,n_r}}\right]^{\pm1}\,.
\end{equation}

The operators of the first pair ($m=1$) are no genuine leader operators since 
they produce the variation with one unit of the parameter $k$ which is not a 
quantum number. Nevertheless, these operators are useful in combination with 
the supersymmetry generators which increase or decrease the value of $k$ in 
a similar way.

The second and third pairs are important because these represent the leader 
operators shifting the principal quantum number $n=2n_r+l$. In terms of the 
quantum numbers $(n,l)$, the second pair performs the shifts $\Delta n=1$ and
$\Delta l=1$ while the shifts given by the third one are $\Delta n=-1$ and 
$\Delta l=1$. We observe that these operators can not be defined in a 
global manner, independent on the values of the parameters 
$(l,n_r)$ or $(n,l)$, since the dependence on $l$ can not be removed 
as long as a differential operator with eigenvalues $l$ does not exist. 

The fourth solution $(m=4)$ yields the unique pair of leader operators which 
can be put in a suitable closed form, depending only on the principal parameter 
of the model, $k$. Indeed, if we proceed as in Ref. \cite{C2} introducing the 
non-differential operator $\hat N$ defined as
\begin{equation}
\hat N R_{k,n,l}=n R_{k,n,l}\,,
\end{equation}
then we observe that the operators $2\,{_4\tilde A}^{(\pm)}_{k,l,n_r}$ 
produce the same radial effects as the operators 
\begin{equation}
A^{(\pm)}_k=\pm2(1-u^2)\frac{d}{du}-u(\hat N+k)
+\frac{\Delta_s+k(k-d)+2d-1}{\hat N+k+1}\,,
\end{equation}
which act in the space of the normalized wave functions as
\begin{equation}
A^{(\pm)}_k R_{k,n,l}Y^{d-1}_{l(\lambda)}=C^{(\pm)}_{k,n,l} R_{k, n\pm 2, l}
Y^{d-1}_{l(\lambda)}\,,
\end{equation}
where
\begin{eqnarray}
C^{(-)}_{k,n,l}&=&
\left[\frac{(n+k)(n-l)}{n+k-2}\right]^{\frac{1}{2}}\\
&\times&\frac{[(n+2k+l-2)(n+2k-l-d)(n+l+d-2)]^{\frac{1}{2}}}{n+k-1}\,, \\
C^{(-)}_{k,n,l}&=&
\left[\frac{(n+k-2)(n-l)}{n+k}\right]^{\frac{1}{2}}\\
&\times&\frac{[(n+2k+l-2)(n+2k-l-d)(n+l+d-2)]^{\frac{1}{2}}}{n+k-1}\,. 
\end{eqnarray}
We note that it is not clear if these operators are adjoint to each other 
with respect to the scalar product (\ref{sc}). This is because it is not 
certain if the coordinate operator commutes with $\hat N$. We hope that further 
investigations should enlighten this delicate point.

Concluding we can say that the results presented here are in accordance 
with our previous results we have obtained for the relativistic 
$(1+1)$-dimensional P\" oschl-Teller problems \cite{C2,C3}. One 
can convince ourselves that the operators shifting the quantum 
number $n$ (of the pairs with $m=2,3$) have similar forms to those of the 
leader operators of the models with $d=1$ \cite{C2}. 
The difference is that for $d>1$ there are many pairs of leader operators  
which could lead to a more complicated dynamical algebra as indicated in 
\cite{B}.


\end{document}